\documentstyle[epsbox]{article}
\huge
\renewcommand{\theequation}{\arabic{section}.\arabic{equation}} 
\def\setzero{\setcounter{equation}{0}}

\newcounter{eqalph}
\def\bph{\setcounter{eqalph}{\value{equation}}
   \addtocounter{eqalph}{1}
   \setcounter{equation}{0}
   \renewcommand{\theequation}{\arabic{section}.\arabic{eqalph}\alph{equation}}}
\def\eph{\setcounter{equation}{\value{eqalph}}
   \renewcommand{\theequation}{\arabic{section}.\arabic{equation}}
\par\noindent}

\begin{document}

\baselineskip 18pt

\def \sech{{\rm sech}}
\def \tanh{{\rm tanh}}
\def \cn{{\rm cn}}
\def \sn{{\rm sn}}
\def\bm#1{\mbox{\boldmath $#1$}}
\title{
 Coupled Nonlinear Schr\"{o}dinger equation
\\
and
\\ 
Toda equation
\\
(the Root of Integrability)} 
\date{\today}
\author{ Masato {\sc Hisakado}  
\\
\bigskip
\\
{\small\it Graduate School of Mathematical  Sciences,}\\
{\small\it University of Tokyo,}\\
{\small\it  Megro-ku, Tokyo 153, Japan}}
\maketitle

\vspace{20 mm}

Abstract

We consider the relation between the discrete coupled nonlinear Schr\"{o}dinger equation and  Toda equation.
Introducing  complex times 
we can show  the intergability 
of the discrete coupled nonlinear Schr\"{o}dinger  equation.
In the same way we can show  the integrability in  coupled  case 
of dark and bright equations.
Using this method we  obtain several integrable equations.

\vfill
\newpage

\section{Introduction}

The propagation of soliton envelopes in nonlinear 
optical media has been predicted and demonstrated experimentally.\cite{ht}
,\cite{m}
This prediction arises from a reduction  the Maxwell equations 
which govern the electro magnetic fields in the medium  to a single,
 completely integrable, partial differential equation.
The well-studied example is the nonlinear Schr\"{o}dinger (NLS) equation:
\begin{equation}
{\rm i}\frac{\partial \phi}{\partial t}
+\frac{\partial^{2} \phi}{\partial  x^{2}}
+2|\phi|^{2}\phi=0.
\label{nls}
\end{equation}
This equation describes the wave propagation of 
picosecond pulse envelopes $\phi(x,t)$ in a 
lossless single mode fiber.
The NLS equation (\ref{nls}) is the one of the completely integrable system.\cite{zs}
Recently the interactions  among  several  modes 
are studied.\cite{my}-\cite{h1}
In general the coupled mode approach
still permits description of the pulse propagation in a multi-mode 
waveguide by means of vector version of (\ref{nls}).
Although these systems of equations are no longer integrable
except for the special parameters,
 one may obtain quantitative information  about the 
pulse propagation restoring  to numerical  and  perturbative methods.
Physically interesting situations   that can be described
by coupled  NLS equations include  two parallel waveguides coupled through
field overlap.
The study of the propagation of optical soliton in multi-mode nonlinear 
couplers is important in the view point of their possible 
applications in  technology.
From the detailed theoretical investigations, several integrable 
 coupled NLS equations  possessing soliton have been introduced
 for the special parameters.

In this paper  we consider the discrete coupled NLS equation.
This equation is integrable
and  $N$-soliton solution is obtained by the Hirota method.\cite{o}
In the continuum limit this equation becomes  the coupled NLS equation.
This equation is embedded in the 2-dimensional Toda 
equation.\cite{ut}
There,  two times $t$ and $\bar{t}$ are  complex conjugate.
From the relation to the Toda equation 
we study the integrability of the discrete coupled NLS equation.
 Using this method we  
show the integrability of the new  equations.

This paper is organized as follows.
In \S 2
we consider the  discrete coupled NLS equation.
In  any coupled  case of the bright and dark equations 
we show the integrability.
In \S 3 
we present the integrable new equation 
using the method in \S  2.
In \S 4
we discuss  this systems from a viewpoint of the conservation laws.
The last section is  devoted to the concluding remarks.

\setzero
\section{Discrete coupled nonlinear Schr\"{o}dinger
 and Toda equation}

Let $\phi_{n}^{(j)}(t)$, $j=1,2,\cdots,l,$ denote 
$l$ component dynamical variables. 
We consider the discrete coupled 
nonlinear Schr\"{o}dinger (DCNS) equation.\cite{o}
\begin{equation}
\frac{\partial \phi^{(j)}_{n}}{\partial t_{1}}
-{\rm i}(\phi_{n-1}^{(j)}+\phi_{n+1}^{(j)})
(1+\kappa \sum_{k=1}^{l}|\phi_{n}^{(k)}|^{2})
+2{\rm  i}\phi_{n}^{(j)}=0,
\label{dcns1}
\end{equation}
where $t_{1}$ is the time and $\kappa$ is a constant.

We introduce the new time $t_{2}$ and 
 assume that $\phi_{n}^{(j)}(t_{1},t_{2})$  satisfies  the following  equation: 
\begin{equation}
\frac{\partial \phi^{(j)}_{n}}{\partial t_{2}}
+(\phi_{n-1}^{(j)}-\phi_{n+1}^{(j)})
(1+\kappa \sum_{k=1}^{l}|\phi_{n}^{(k)}|^{2})
=0
\label{dcns2}
\end{equation}
This  means a restriction that $\phi_{n}^{(j)}(t_{1},t_{2})$ is the solution 
of (\ref{dcns1}) and (\ref{dcns2}).
The relation between (\ref{dcns1}) and (\ref{dcns2})
is ``dual''.
We call (\ref{dcns2}) the coupled modified Volterra (CMV) equation.

We find it convenient to use  complex times $t$ and $\bar{t}$
which is related to 
$t_{1}$ and $t_{2}$.
\begin{equation}
t=t_{2}+{\rm i}t_{1},\;\;\;\bar{t}
=t_{2}-{\rm i}t_{1}.
\end{equation}
In terms of the  complex times,  we rewrite (\ref{dcns1}) and (\ref{dcns2}):
\bph
\begin{equation}
\frac{\partial \phi_{n}^{(j)}}{\partial t}
=
\phi_{n+1}^{(j)}(1+\kappa\sum_{k=1}^{l}|\phi_{n}^{(k)}|^{2})
-\phi_{n}^{(j)},
\end{equation}
\begin{equation}
\frac{\partial \phi_{n}^{(j)}}{\partial \bar{t}}
=
-\phi_{n-1}^{(j)}(1+\kappa\sum_{k=1}^{l}|\phi_{n}^{(k)}|^{2})
+\phi_{n}^{(j)},
\end{equation}
\begin{equation}
\frac{\partial \bar{\phi}_{n}^{(j)}}{\partial t}
=
-\bar{\phi}_{n-1}^{(j)}(1+\kappa\sum_{k=1}^{l}|\phi_{n}^{(k)}|^{2})
+\bar{\phi}_{n}^{(j)},
\end{equation}
\begin{equation}
\frac{\partial \bar{\phi}_{n}^{(j)}}{\partial \bar{t}}
=
\bar{\phi}_{n+1}^{(j)}(1+\kappa\sum_{k=1}^{l}|\phi_{n}^{(k)}|^{2})
-\bar{\phi}_{n}^{(j)}.
\end{equation}
\eph
Here we introduce new dependent variables:
\begin{equation}
a_{n}^{(j)}=\kappa|\phi_{n}^{(j)}|^{2},
\;\;\;\;\;
b_{n}^{(j)}=\kappa\phi_{n}^{(j)}\bar{\phi}_{n-1}^{(j)},
\;\;\;\;\;
\bar{b}_{n}^{(j)}=\kappa\bar{\phi}_{n}^{(j)}\phi_{n-1}^{(j)}.
\end{equation}
The physical meanings of $a_{n}^{(j)}$ and $b_{n}^{(j)}$
are the amplitude and the momentum.
We rewrite (2.4) using these variables
\bph
\begin{equation}
\frac{\partial a_{n}^{(j)}}{\partial t}
=
(b_{n+1}^{(j)}-b_{n}^{(j)})(1+\kappa\sum_{k=1}^{l}
a_{n}^{(k)}),
\end{equation}
\begin{equation}
\frac{\partial a_{n}^{(j)}}{\partial \bar{t}}
=
(\bar{b}_{n+1}^{(j)}-\bar{b}_{n}^{(j)})(1+\kappa\sum_{k=1}^{l}
a_{n}^{(k)}),
\end{equation}
\begin{equation}
\frac{\partial b_{n}^{(j)}}{\partial \bar{t}}
=
-a_{n-1}^{(j)}(1+\kappa\sum_{k=1}^{l}a_{n}^{(k)}) 
+a_{n}^{(j)}(1+\kappa\sum_{k=1}^{l}a_{n-1}^{(k)}),
\end{equation}
\begin{equation}
\frac{\partial \bar{b}_{n}^{(j)}}{\partial t}
=
-a_{n-1}^{(j)}(1+\kappa\sum_{k=1}^{l}a_{n}^{(k)}) 
+a_{n}^{(j)}(1+\kappa\sum_{k=1}^{l}a_{n-1}^{(k)}),
\end{equation}
\eph

It is remarkable that these equations are the 2-dimensional Toda lattice equation:
\bph
\begin{equation}
\frac{\partial a_{n}}{\partial t}
=
(b_{n+1}-b_{n})a_{n},
\;\;\;\;
\frac{\partial a_{n}}{\partial \bar{t}}
=
(\bar{b}_{n+1}-\bar{b}_{n})a_{n},
\label{tl1}
\end{equation}
and
\begin{equation}
\frac{\partial b_{n}}{\partial \bar{t}}
=
\frac{\partial \bar{b}_{n}}{\partial t}
=
a_{n}-a_{n-1},
\label{tl2}
\end{equation}
\eph
where
\begin{equation}
a_{n}=1+\sum_{j=1}^{l}a_{n}^{(j)},
\;\;\;\;\;
b_{n}=\sum_{j=1}^{l}b_{n}^{(j)},\;\;\;
\bar{b}_{n}=\sum_{j=1}^{l}\bar{b}_{n}^{(j)}
\end{equation}
We note that  $a_{n}$ and $b_{n}$ 
are the sum of the amplitudes and the sum of the momentum 
over   all components.
It means that
DCNS and CMV equations are  the variants of the 2-dimensional Toda lattice equation.
However (\ref{tl1}) and (\ref{tl2}) are not regular
2-dimensional Toda lattice.
This equation contains complex conjugate
in $t$ ($\bar{t}$) and $b_{n}$ ($\bar{b}_{n}$).
The Toda equation   has the $N$-soliton solution.
Then both DCNS and CMV equations have 
solutions corresponding to the solutions of the  Toda lattice. 

Hereafter  we call the  two equations are ``dual'' when 
from these equations we can construct the Toda lattice equation.

Next we consider the coupled case of the  bright and  dark equations:
\begin{eqnarray}
& &
\frac{\partial \phi_{n}^{(j)}}{\partial t_{1}}
-{\rm i}(\phi_{n-1}^{(j)}+\phi_{n+1}^{(j)})
(1+\kappa \sum_{j=1}^{l}|\phi_{n}^{(j)}|^{2}
-\kappa\sum_{j=l+1}^{m}|\phi_{n}^{(j)}|^2)
+2{\rm i}\phi_{n}^{(j)}=0,
\nonumber \\
& &
j=1,2,\cdots,l
\end{eqnarray}
\begin{eqnarray}
& &
\frac{\partial \phi_{n}^{(j)}}{\partial t_{1}}
+{\rm i}(\phi_{n-1}^{(j)}+\phi_{n+1}^{(j)})
(1+\kappa \sum_{j=1}^{l}|\phi_{n}^{(j)}|^{2}
-\kappa\sum_{j=l+1}^{m}|\phi_{n}^{(j)}|^2)
+2{\rm i}\phi_{n}^{(j)}=0.
\nonumber \\
& &
j=l+1,l+2,\cdots,m
\end{eqnarray}

Among  the $m$-components equations,
$l$ equations are bright-type and $(m-l)$ equations are dark-type.
We consider also the ``dual'' equations:
\begin{eqnarray}
& &\frac{\partial \phi_{(j)}}{\partial t_{2}}
+(\phi_{n-1}^{(j)}-\phi_{n+1}^{(j)})
(1+\kappa \sum_{j=1}^{l}|\phi_{n}^{(j)}|^{2}
-\kappa\sum_{j=l+1}^{m}|\phi_{n}^{(j)}|^2)
=0,
\nonumber \\
& &
j=1,2,\cdots,l
\end{eqnarray}
\begin{eqnarray}
& &
\frac{\partial \phi_{(j)}}{\partial t_{2}}
+(-\phi_{n-1}^{(j)}+\phi_{n+1}^{(j)})
(1+\kappa \sum_{j=1}^{l}|\phi_{n}^{(j)}|^{2}
-\kappa\sum_{j=l+1}^{m}|\phi_{n}^{(j)}|^2)
=0,
\nonumber \\
& & j=l+1,l+2,\cdots,m
\end{eqnarray}

In the same way introducing  the complex times
we can cast  
these equations as  the Toda lattice equation (\ref{tl1}) and (\ref{tl2}).
The only difference is the definition of $a_{n}$:
\begin{equation}
a_{n}=1+\sum_{j=1}^{l}a_{n}^{(j)}
-\sum_{j=l+1}^{m}a_{n}^{(j)}.
\end{equation}
That is 
$a_{n}$ is the sum of the amplitudes
 but for the dark-type 
the sign is ``-''.
Usual one dark soliton is described 
as $A-B\sech^{2} X$.
Then from the view point of the Toda equation 
the dark and bright soliton are same.

\setzero
\section{New Integrable Equations}

Here we consider new coupled equations:
\begin{equation}
\frac{\partial \phi_{n}^{(j)}}{\partial t_{1}}
-{\rm i}(\phi_{n-1}^{(j)}+\phi_{n+1}^{(j)})
(1+\kappa \sum_{k, l}g_{kl}\phi_{n}^{(k)}\bar{\phi}_{n}^{(l)})
+2{\rm i}\phi_{n}^{(j)}=0.
\label{n1}
\end{equation}
where 
\begin{equation}
g_{ij}=g_{ji},\;\;\;\;
g_{ij}=0 \;\;\;{\rm or} \;\;\;1.
\end{equation}
As in  the previous section 
we introduce 
the ``dual '' equation 
\begin{equation}
\frac{\partial \phi_{(j)}}{\partial t_{2}}
+(\phi_{n-1}^{(j)}-\phi_{n+1}^{(j)})
(1+\kappa \sum_{k,l}g_{kl}\phi_{n}^{(k)}\bar{\phi}_{n}^{(l)})
=0.
\label{n2}
\end{equation}

Moreover we define new dependent variables:
\begin{equation}
a_{n}^{(ij)}=\kappa\phi_{n}^{(i)}\bar{\phi}_{n}^{(j)},
\;\;\;
b_{n}^{(ij)}=\kappa\phi_{n}^{(i)}\bar{\phi}_{n-1}^{(j)},
\;\;\;
\bar{b}_{n}^{(ij)}=\kappa\bar{\phi}_{n}^{(i)}\phi_{n-1}^{(j)}. 
\end{equation}

In the same way introducing the complex times,
 we can obtain the Toda lattice equation (\ref{tl1}) and (\ref{tl2})
for the variables:
\begin{equation}
a_{n}=1+\sum_{i,j}g_{ij}a_{n}^{(ij)},
\;\;\;
b_{n}=\sum_{i,j}g_{ij}b_{n}^{(ij)},
\;\;\;
\bar{b}_{n}=\sum_{i,j}g_{ij}\bar{b}_{n}^{(ij)}.
\end{equation}
From these results we can see that (\ref{n1}) and (\ref{n2})
are integrable.

If we set $g_{ij}=\delta_{ij}$ then 
(\ref{n1}) and (\ref{n2}) become
DCNS and CMV equations respectively.
If we set $g_{ij}=1$ then 
(\ref{n1}) and (\ref{n2}) 
can be reduced 
to the (no coupled) discrete nonlinear Schr\"{o}dinger (DNLS)
and the (no coupled) modified Volterra (MV) equations.
As  examples 
we  obtain the following new integrable equations:

i)
\begin{equation} 
\frac{\partial \phi_{n}^{(j)}}{\partial t_{1}}
-{\rm i}(\phi_{n-1}^{(j)}+\phi_{n+1}^{(j)})
(1+\kappa \sum_{k\neq l}\phi_{n}^{(k)}\bar{\phi}_{n}^{(l)})
+2{\rm i}\phi_{n}^{(j)}=0.
\label{ne1}
\end{equation}

ii)
\begin{eqnarray} 
& & \frac{\partial \phi_{(j)}}{\partial t_{1}}
-{\rm i}(\phi_{n-1}^{(j)}+\phi_{n+1}^{(j)})
(1+\kappa \sum_{k}^{l}
\phi_{n}^{(k)}\bar{\phi}_{n}^{(k+1)})
+2{\rm i}\phi_{n}^{(j)}=0,
\nonumber \\
& & \phi_{n}^{(k+l)}=\phi_{n}^{(k)},\;\;\;\;
\bar{\phi}_{n}^{(k+l)}=\bar{\phi}_{n}^{(k)}.
\label{ne2}
\end{eqnarray}
In the case (i) or (ii) 
the soliton solution  in a line must run together 
 with the soliton in the other line. 
On the other hands in the case DCNS (\ref{dcns1}) equation
the soliton solution  can  run independently.
If the difference of the phases between $j$ and $k$ lines is
$-\pi/2<\theta^{ij}<\pi/2$,
 the amplitude becomes $a_{n}^{(jk)}<0$ and the 
solution is ``dark''. 
(\ref{n1}) and (\ref{n2}) may be integrable in the continuum  limit.

The discrete Hirota equation reads as:\cite{n}
\begin{equation}
\frac{\partial \phi_{n}}{\partial t_{1}}
-{\rm i}\alpha(\phi_{n-1}+\phi_{n+1})
(1+\kappa |\phi_{n}|^{2})
+\beta(\phi_{n-1}-\phi_{n+1})
(1+\kappa |\phi_{n}|^{2})
+2i\alpha\phi_{n}=0
\label{he}
\end{equation}
This equation is a hybrid of 
the discrete nonlinear Schr\"{o}dinger (DNLS) equation 
and the modified Volterra  (MV) equation.
In the continuous limit (\ref{he}) becomes
\begin{equation}
{\rm i}\frac{\partial \phi}{\partial t}
+k_{1} \frac{\partial^{2} \phi}{\partial x^{2}}
+k_{2}|\phi|^{2}\phi
+
{\rm i}[k_{3}\frac{\partial^{3} \phi}{\partial x^{3}}
+k_{4}|\phi|^{2}\frac{\partial \phi }{\partial x}]=0.
\label{he1}
\end{equation}
where $k_{i}$ is the arbitrary parameters.
(\ref{he1}) contains several  generalized (continuous) NLS equations.
As pulse lengths become comparable  to the 
wavelength, NLS equation is not adequate, 
as  the additional effects  must be considered.
For these cases (\ref{he1})  is useful.\cite{hh}

We consider  the coupled version of the discrete 
Hirota (DCH)  equation:
\begin{eqnarray}
& &
\frac{\partial \phi_{(j)}}{\partial t_{1}}
-{\rm i}\alpha(\phi_{n-1}^{(j)}+\phi_{n+1}^{(j)})
(1+\kappa \sum_{k=1}^{l}|\phi_{n}^{(k)}|^{2})
+\beta(-\phi_{n-1}^{(j)}+\phi_{n+1}^{(j)})
(1+\kappa \sum_{k=1}^{l}|\phi_{n}^{(k)}|^{2})
\nonumber
\\
& &
+2{\rm i}\alpha\phi_{n}^{(j)}=0.
\label{n3}
\end{eqnarray}
In the same way we introduce the
``dual'' equation:
\begin{eqnarray}
& &
\frac{\partial \phi^{(j)}_{n}}{\partial t_{2}}
-{\rm i}\beta(\phi_{n-1}^{(j)}+\phi_{n+1}^{(j)})
(1+\kappa \sum_{k=1}^{l}|\phi_{n}^{(k)}|^{2})
+\alpha(\phi_{n-1}^{(j)}-\phi_{n+1}^{(j)})
(1+\kappa \sum_{k=1}^{l}|\phi_{n}^{(k)}|^{2})
\nonumber \\
& &
+2{\rm i}\beta\phi_{n}^{(j)}
=0.
\label{n4}
\end{eqnarray}

Notice that the ``dual '' equation also becomes the
(DCH) equation,
with  the parameters $\alpha$ and $\beta$ being  exchanged.

We introduce the following new independent variables:
\begin{equation}
a_{n}^{(j)}=\kappa|\phi_{n}^{(j)}|^{2},
\;\;\;
b_{n}^{(j)}=\kappa(\alpha+\beta {\rm i})\phi_{n}^{(j)}
\bar{\phi}_{n-1}^{(j)},
\;\;\;
\bar{b}_{n}^{(j)}=\kappa(\alpha-\beta {\rm i})\bar{\phi}_{n}^{(j)}
\phi_{n-1}^{(j)}.
\end{equation}
 
Then we can obtain the Toda lattice equation (\ref{tl1}) and (\ref{tl2}) 
for  the variables:
\begin{equation}
a_{n}=(\alpha^{2}+\beta^{2})(1+\sum_{j=1}^{l}
a_{n}^{(j)}),\;\;\;
b_{n}=\sum_{j=1}^{(l)}b_{n}^{(j)},
\;\;\;
\bar{b}_{n}=\sum_{j=1}^{(l)}\bar{b}_{n}^{(j)}.
\end{equation}

Form theses results (\ref{n3}) and (\ref{n4}) 
are seen to be  integrable.
These equations are also the variants of the 
Toda lattice equation.
In the case that the bright and dark equations 
are coupled,
such  equations are also integrable as 
discussed in the previous section.

Here we change $\alpha+{\rm i}\beta $ to $ r {\rm e}^{{\rm i}\theta}$.
If we set
$
\tilde{\phi}_{n}^{(j)}={\rm e}^{{\rm i}\theta n+2{\rm i}(r-\alpha)t_{1}}\phi_{n}^{(j)},
$
then $\tilde{\phi}_{n}^{(j)}$ satisfy the 
DCNS equation.
In this meaning DCH (\ref{n3}) equation is the same as 
the DCNS  (\ref{dcns1}) equation. 
But in the continuum  limit
 there are diffrernces.
In the continuum  limit (\ref{n3}) 
becomes  generalized coupled NLS equations:
\begin{eqnarray}
& &
{\rm i}\frac{\partial \phi^{(j)}}{\partial t}
+k_{1} \frac{\partial^{2} \phi^{(j)}}{\partial x^{2}}
+k_{2}(\sum_{l}|\phi^{(l)}|^{2})\phi^{(j)}
+
{\rm i}k_{3}\frac{\partial^{3} \phi^{(j)}}{\partial x^{3}}
\nonumber \\
& &
+{\rm i}k_{4}(\sum_{l}|\phi^{(l)}|^{2})\frac{\partial \phi^{(j)} }{\partial x}=0,
\label{he2}
\end{eqnarray}
while (\ref{he2}) contains nonlinear  derivative term. 
\setzero
\section{Conservation Laws}

The 2-dimensional Toda lattice system
(\ref{tl1}) and (\ref{tl2})
has infinite number of conserved quantities.
The   conserved density are obtained 
from the Lax operator. \cite{m3}
\begin{eqnarray}
D_{1}^{T}&=&b_{n},\;\;\;\bar{D}_{1}^{T}=\bar{b}_{n}
\nonumber \\
D_{2}^{T}&=&\frac{1}{2}b_{n}^{2}+\frac{a_{n}}{1-a_{n}}b_{n}b_{n-1}
,\;\;\;
\bar{D}_{2}^{T}=\frac{1}{2}\bar{b}_{n}^{2}+\frac{a_{n}}{1-a_{n}}
\bar{b}_{n}\bar{b}_{n-1},
\nonumber \\
D_{3}^{T}&=&\frac{1}{3}b_{n}^{3}+\frac{a_{n}}{1-a_{n}}
 \frac{a_{n-1}}{1-a_{n-1}}b_{n}b_{n-1}b_{n-2}
\nonumber \\
& &
 +\frac{a_{n}}{1-a_{n}}b_{n}b_{n-1}(b_{n}+b_{n-1}),
\nonumber \\
\bar{D}_{3}^{T}&=&\frac{1}{3}\bar{b}_{n}^{3}+\frac{a_{n}}{1-a_{n}}
 \frac{a_{n-1}}{1-a_{n-1}}\bar{b}_{n}\bar{b}_{n-1}\bar{b}_{n-2}
\nonumber \\
& &
 +\frac{a_{n}}{1-a_{n}}\bar{b}_{n}\bar{b}_{n-1}(
\bar{b}_{n}+\bar{b}_{n-1}),
\label{tc}
\end{eqnarray}
On the other hand
 the conserved densities  of the DNLS  and MV equation 
 are 
obtained from the  Abolowitz and Ladik (AL) system:\cite{al}
\begin{eqnarray}
D_{1}^{AL}&=&\phi_{n}\bar{\phi}_{n-1}
,\;\;\;\bar{D}_{1}^{AL}=\bar{\phi}_{n}\phi_{n-1}
\nonumber \\
D_{2}^{AL}&=&\frac{1}{2}(\phi_{n}\bar{\phi}_{n-1})^{2}
+\phi_{n+1}\bar{\phi}_{n}(1+|\phi_{n}^{j}|^{2})
\nonumber \\
\bar{D}_{2}^{AL}&=&\frac{1}{2}(\bar{\phi}_{n+1}\phi_{n-1})^{2}
+\bar{\phi}_{n+1}\phi_{n-1}(1+|\phi_{n}|^{2}),
\nonumber \\
D_{3}^{AL}&=&\frac{1}{3}(\phi_{n}^{(j)}\bar{\phi}_{n-1})^{3}
  +\phi_{n+1}\bar{\phi}_{n-2}(1+|\phi_{n}|^{2})(1+|\phi_{n-1}|^{2})
\nonumber \\
& &
+\phi_{n+1}\bar{\phi}_{n-1}(\phi_{n+1}\bar{\phi}_{n}
+\phi_{n}\bar{\phi}_{n-1})(1+|\phi_{n}|^{2}),
\nonumber \\
\bar{D}_{3}^{AL}&=&\frac{1}{3}(\bar{\phi}_{n}^{(j)}\phi_{n-1})^{3}
  +\bar{\phi}_{n+1}\phi_{n-2}(1+|\phi_{n}|^{2})(1+|\phi_{n-1}|^{2})
\nonumber \\
& &
+\bar{\phi}_{n+1}\phi_{n-1}(\bar{\phi}_{n+1}\phi_{n}
+\bar{\phi}_{n}\phi_{n-1})(1+|\phi_{n}|^{2}).
\end{eqnarray}
The conserved densities  of the Toda equation 
agree with those of the AL system.
For the multi component cases 
the number of the conserved densities  
is the same as the one component case.\cite{h1}
Then, the conserved densities  of the Toda equation 
agree  with those the multi component AL systems.
From these  results  we can obtain the conserved density for the
DCNS,  CMV  and 
new integrable equations using  (\ref{tc}).  
\setzero
\section{Concluding Remarks}

In this paper we have studied  the 
discrete coupled nonlinear Schr\"{o}dinger (DCNS) and 
coupled modified Volterra (CMV) equations 
from the view point of the 2-dimensional Toda  equation.
Introducing the complex time, coupling the DCNS and CMV
equations we have obtained  the Toda equation.
The DCNS and CMV equation are equivalent 
to the Toda system.
This  can be seen from the explicit transformations and conservation laws.
From this point of view 
when the bright and the dark equation are coupled,
 the DCNS equation is also integrable.
In the Toda equation 
the dark and bright soliton 
are equivalent. 
Using this method we can present 
several new equations.
 All these equations are embedded 
in the 2-dimensional Toda equation 
of which time variable $t$ and $\bar{t}$ are  complex conjugate.
In the continuum  limit these new equations may be integrable.

\clearpage
\end{document}